# Semiclassical approach to tunneling through a double barrier structure

J. C. Martinez[*] & E. Polatdemir[*]

Natural Sciences Academic Group
National Institute of Education
Nanyang Technological University
1 Nanyang Walk, Singapore 637616

We consider a symmetric double barrier heterostructure enclosing a well and propose a solution for the transmission problem using a generalized WKB approach which accounts for the amplitude suppression and phase shift due to the barriers. This approach allows us to address both off-resonance and resonance cases and, in the latter case, verify the 'coherent destruction of tunneling.'

____________________

[*]email: jcmartin@nie.edu.sg
[*]email: hffb39@mail.nie.edu.sg

## I. Introduction

In recent years barrier transmission problems have become the target of considerable attention brought on by the discoveries of new experimental and materials-preparation techniques and theoretical investigations [1]. For instance, numerical studies of double-barrier resonant tunneling in which the quantum well is subject to a periodic modulation indicate that the system possesses a far richer spectrum than previously suspected [2]. Earlier Buttiker and Landauer had discussed the transport of electrons through a harmonically oscillating barrier and they showed that its analysis can provide important new insights into the tunneling process [3]. As a consequence, the need for better approximation methods for calculating transmission probabilities has also intensified [4]. Although the WKB approximation method is one of the oldest available, it remains a reliable standard method for computations, in part, because it is not a perturbation expansion and because it remains accurate provided the particle momentum is constant over many wavelengths. Nevertheless, the WKB method always breaks down at the classical turning points (where particle energy equals the potential) so to deal with this limitation the phases and amplitudes of the WKB wave functions on either side of a turning point are joined together by connections formulas that are conventionally derived under the condition of short wavelengths. During the past five years, efforts to improve on these formulas have centered on the determination of suitable energy-dependent expressions for the phases and amplitudes of the wave functions at the turning points, in marked contrast with the constant values that had been in use for a long time. Indeed the suppression of the wave function at the classically forbidden side of a turning point is known to occur when the short wavelength limit is not satisfied. [5]. And there is no a priori reason to assume that the amplitudes and phases remain energy independent either. This new program has been successfully implemented by a number of workers. One way to carry this out has been to introduce real amplitude factors $N$ and $\overline{N}$ and corresponding real phases $\phi$ and $\overline{\phi}$ on the two sides of a turning point and to generalize the usual WKB connections formulas to

$$\frac{2}{\sqrt{p(x)}}\cos\left(\frac{1}{\hbar}\left|\int_x^x p(x)dx\right|-\frac{\phi}{2}\right) \leftrightarrow \frac{N}{\sqrt{|p(x)|}}\exp\left(-\frac{1}{\hbar}\left|\int_x^x p(x)dx\right|\right)$$

$$\frac{1}{\sqrt{p(x)}}\cos\left(\frac{1}{\hbar}\left|\int_x^x p(x)dx\right|-\frac{\overline{\phi}}{2}\right) \leftrightarrow \frac{\overline{N}}{\sqrt{|p(x)|}}\exp\left(\frac{1}{\hbar}\left|\int_x^x p(x)dx\right|\right),$$

(1)

subject to a consistency condition $N\overline{N} = \sin \frac{1}{2}(\phi - \overline{\phi})$ [6]. Here the particle kinetic energy is $E$ and $\hbar p(x) = [2m(E-V(x))]^{1/2}$ represents its classical momentum in an allowed region with the corresponding expression $\hbar \overline{p}(x) = [2m(V(x)-E)]^{1/2}$ in a classically forbidden region. One now proceeds along the lines prescribed by the standard WKB theory. To recover the conventional WKB theory, one sets $N = \overline{N} = 1$ and $\phi = -\overline{\phi} = \pi/2$ [7]. However, Moritz has shown that $\overline{N}$ is almost never unity in many examples he had looked at and is even zero in the semiclassical limit for the inverted oscillator [5]. Applications of this procedure for suitable choices of $N$, $\overline{N}$, $\phi$ and $\overline{\phi}$ have meet with some success for a number of barrier types [8].

Table I: Phase and weight factors at each turning point.

| | Phase | Weight |
|---|---|---|
| Allowed → allowed | $e^{-i\phi}$ | 1 |
| Forbidden→forbidden | $-e^{-i(\phi-\overline{\phi})/2}$ | $N/2\overline{N}$ |
| Forbidden→allowed | $ie^{-i\phi/2}$ | $N$ |
| Allowed→forbidden | $e^{-i\phi/2}$ | $N$ |

By virtue of the fact that the WKB approach can be reproduced by a semi-classical evaluation of the corresponding complex-time path integral [9], a second similarly inspired and equivalent procedure seeks to assign weight factors and phases at the turning points according to Table I [9 - 11]. The same consistency condition holds here also. Again for this semiclassical approach, no specific prescription has yet been given for $N$, $\overline{N}$, $\phi$, and $\overline{\phi}$ although the choice $N = \overline{N} = 1$ and $\phi = -\overline{\phi} = \pi/4$ has been advocated by Aoyama and Harano [11]. The entries displayed in Table I were obtained by studying the complex time paths taken by a particle crossing a barrier approximated as a linear potential at the vicinity of a turning point. Unlike the first way described above, this second way does not make explicit use of the wave functions. Instead, propagation factors are assigned between turning points $a$ and $b$ according to the following rules:

(I) if $a$ and $b$ are in a classically allowed region append the factor

$$\exp\left( i \int_a^b p(x)dx \right); \tag{2a}$$

(II) if $a$ and $b$ are in a classically forbidden region append the factor

$$\exp\left( -\int_a^b \overline{p}(x)dx \right). \tag{2b}$$

In energy space the semi-classical propagator between points $x$ and $x'$ in a classically accessible regions is given by [9, 12]

$$D(x', x; E) = \frac{m}{2\pi\hbar\sqrt{p(x)p(x')}} \sum_{\text{paths}} \prod_i f_i, \tag{3}$$

where the sum is over all fixed energy paths connecting the end points and the factors $f_i$ are the amplitude factors, phases and propagation factors mentioned above. The idea of following paths instead of writing out wave functions has a more intuitive appeal than a wave function approach.

Although, as noted already, no unique prescription for choosing the amplitude factors and phases has been found, it is known that for a particle with energy $E$ incident on a rectangular barrier of height $V_0$ ($> E$) and separation $d$ the choice

$$N = 2\overline{N} = 2\sqrt{\frac{kq}{k^2 + q^2}}, \quad \phi = -\overline{\phi} = 2\arctan\left(\frac{q}{k}\right), \tag{4}$$

where $k = \sqrt{2mE}/\hbar$, $q = \sqrt{2m(V_0 - E)}/\hbar$ yields the exact quantum-mechanical result [8]. Our goal here is to apply these results to the problem of particle transmission through a double barrier in the case that a harmonically oscillating well is present [13]. We will, of course, look first at the double barrier without the oscillating well. Besides the computational tractability offered, there are several reasons for studying this system: (1) thus far the best WKB calculations for the double barrier alone have not successfully reproduced the exact quantum mechanical results so it is worthwhile inquiring whether the semiclassical approach can be more successful [7, 14]; (2) both the WKB and the semiclassical methods are usually applied to time-independent problems whereas the present problem offers an avenue into a time-dependent situation [15]; (3) as an offshoot of our work, another approach (essentially nonperturbative in the modulation amplitude) to the exciting phenomenon of the quenching of resonant transmission through an oscillating well has become available; (4) the double barrier is the object of intense experimental and theoretical study at the present time [1]; by contrast, studies on general approximations for potential transmission and reflection problems center on the comparison of results from different theoretical approaches. Ultimately, though, we hope that a successful description of the tunneling process for this simple system would give us confidence to study, within the semiclassical approach, other cases that are less amenable to analytical calculation.

## II. Transmission through a Static Double Barrier

Let us now consider the transmission of electrons with energy $E$ through a double barrier structure of height $V_0$ ($> E$) as shown in Fig. 1. The potential is first divided into five regions labeled 0, 1, 2, 3, and 4. We consider first the case that there is no oscillating well sandwiched between the static barriers. To proceed efficiently with the calculation, it will be convenient to follow a procedure given by Holstein and Swift [14].

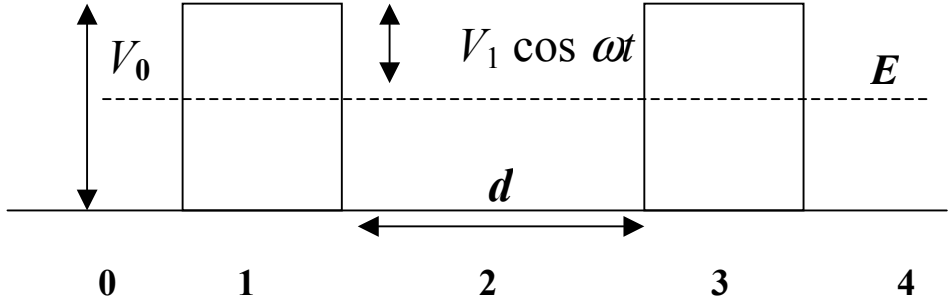

FIG 1  Transmission through a double barrier structure with an oscillating quantum well between the static barriers.

Let $T_{ij}^{(\pm)}$ denote the amplitude for the particle to enter the region $i$ from the left (+) or right (-) and to eventually emerge into region $j$, via trajectories that are all confined to region $i$.  Following Eq. (3) and the prescription of Table I as well as the propagation factors of Eqs. (2),  we write the following expressions for the $T$'s,

$$T_{12}^{(+)} = T_{34}^{(+)} = Ne^{-i\phi/2}\left(e^{-\theta}\sum_{n=0}^{\infty}\left(-\frac{N}{2\overline{N}}e^{-i\phi}e^{-\theta}\right)^{2n}\right)iNe^{-i\phi/2} = \frac{iN^2 e^{-i\phi}e^{-\theta}}{1-\left(\dfrac{N}{2\overline{N}}\right)^2 e^{-2i\phi}e^{-2\theta}},$$

(5)

$$T_{12}^{(-)} = T_{32}^{(+)} = -\frac{N}{2\overline{N}}e^{-i\phi}e^{-\theta}T_{12}^{(+)},$$  (6)

$$T_{21}^{(+)} = iN\left(e^{-i\phi/2}e^{2iL}e^{-i\phi}\right)\left(\sum_{n=0}^{\infty}\left(e^{-i\phi}e^{iL}\right)^{2n}\right)Ne^{-i\phi/2} = -\frac{1}{2}\frac{N^2 e^{-i\phi}e^{iL}}{\sin(L-\phi)},$$  (7)

$$T_{23}^{(+)} = T_{21}^{(-)} = e^{-i(L-\phi)}T_{21}^{(+)} = e^{-i(L-\phi)}T_{23}^{(-)},$$  (8)

where $L = kd$ and $\theta = \int_{\text{barrier}} q(x)dx$ in accordance with Eqs. [2].  To account for processes whereby particles may re-enter a region once occupied, we introduce the quantity $B_{ij}^{(\pm)}$ which denotes the amplitude for the particle entering region $i$ from the left (+) or right (-) and escaping, its first destination upon leaving region $i$ being the region $j$. Thus

$$B_{23}^{(+)} = T_{23}^{(+)} \left(\frac{1}{Ne^{-i\phi/2}}\right)\left[T_{34}^{(+)} + T_{32}^{(+)}\left(\frac{1}{Ne^{-i\phi/2}}\right)\left(B_{23}^{(-)} + B_{21}^{(-)}\right)\right], \tag{9}$$

$$B_{21}^{(+)} = T_{21}^{(+)} \left(\frac{1}{Ne^{-i\phi/2}}\right) T_{12}^{(-)} \left(\frac{1}{Ne^{-i\phi/2}}\right)\left[B_{23}^{(+)} + B_{21}^{(+)}\right] \tag{10}$$

$$B_{23}^{(-)} = e^{i(L-\phi)} B_{23}^{(+)}, \tag{11}$$

$$B_{21}^{(-)} = e^{-i(L-\phi)} B_{21}^{(+)}. \tag{12}$$

In writing these equations, factors of $1/Ne^{-i\phi/2}$ were inserted to compensate for their having been double-counted at the interfaces between regions in the expressions for the $T$'s and $B$'s.

Solving simultaneously the above set of four equations we obtain

$$B_{23}^{(+)} + B_{21}^{(+)} = \frac{1/(N e^{-i\phi/2})}{\left(T_{23}^{(+)} T_{34}^{(+)}\right)^{-1} + \frac{N}{\overline{N}} e^{-\theta} \frac{e^{i(L-\phi)}}{N^2} + O(e^{-4\theta})}. \tag{13}$$

The full transmission amplitude $T(E)$ can now be computed explicitly from

$$T(E) = T_{12}^{(+)} \frac{1}{N e^{-i\phi/2}} \left(B_{23}^{(+)} + B_{21}^{(+)}\right). \tag{14}$$

For the amplitude factors and phases we choose those of Eq. (4) for which the phase is [8]

$$\sin\phi = \frac{N^2}{2} = \frac{2kq}{k^2+q^2}.  \tag{15}$$

Simplifying we have

$$T(E) = \frac{-i}{\dfrac{e^{i(L-\phi)}}{\sin\phi} - 2\dfrac{\sinh\theta}{\sin\phi}\sin(L-\phi)[\cosh\theta - i\cot\phi\sinh\theta]} + O(e^{-4\theta}) \tag{16}$$

In most instances, the second term in the denominator dominates so in this case $T(E)$ coincides with the exact off-resonance result up to corrections of order $e^{-4\theta}$ [1].

From Eq. (16) a resonance occurs whenever sin $(L-\phi)$ vanishes. In this instance the first term in the denominator gives the entire contribution to $T(E)$. Since at resonance, $1 = \cos(L - \phi) = \cos(L - \phi/2)\cos(\phi/2) + \sin(L - \phi/2)\sin(\phi/2)$, it follows that $\cos(L - \phi/2) = \cos(\phi/2)$ and $\sin(L - \phi/2)\sin(\phi/2)$ from which the resonance condition of Wagner (based on the existence of bound states in the well) follows [13]:

$$1 = \cos L + \tan\frac{\phi}{2}\sin L = \cos L + \frac{q}{k}\sin L \tag{17}$$

Although Eq. (16) works very well outside resonance, we can improve on it and calculate the transmission amplitude at resonance by stipulating that while the amplitude factors and phases for the boundaries between regions 1 and 2 and between regions 2 and 3 are those of Eltschka et al [8] [i.e. Eq. (4)], the corresponding factors and phases for the boundaries between regions 0 and 1 and between regions 3 and 4 should be those of Aoyama and Harano [11] instead, namely, $N = \overline{N} = 1$ and $\phi = -\overline{\phi} = \pi/4$. As stated above there is no unique prescription for choosing $N$, $\overline{N}$, $\phi$, and $\overline{\phi}$ so we are making use of this freedom here. A careful study of the previous results shows that one choice alone for the amplitudes factors cannot reproduce both the off-resonance and resonance transmission probabilities. In the former case the transmission is less than unity while in the latter it can be 100%.

Fortunately, it is not necessary to go through the whole series of calculations already given to obtain the final result. By virtue of the resonance condition $e^{i(L-\phi)} = 1$, we may proceed more directly as follows. First we observe that the oscillations in the classically accessible region 2 involve the factor $e^{iL}\left[1 + \left(e^{-i\phi}e^{iL}\right)^2 + ....\right]$, which does not converge at resonance. Hence we must modify this by including contributions from tunneling into regions 1 and 3. Thus it is clear that tunneling into either region requires that we modify $e^{iL}e^{-i\phi}$ into

$$e^{iL}\left\{e^{-i\phi} + Ne^{-i\phi/2}e^{-\theta}\frac{i}{2}e^{-\theta}iNe^{-i\phi/2}\right\} = e^{i(L-\phi)}\left[1 - \frac{1}{2}N^2e^{-2\theta}\right]. \qquad (18)$$

This is correct up to terms of order $e^{-4\theta}$. The divergent oscillation factor is now replaced by

$$e^{iL}\left\{1 + \left[e^{i(L-\phi)}\left(1 - \frac{1}{2}N^2e^{-2\theta}\right)\right]^2 + ....\right\} \approx \frac{e^{iL}}{1 - e^{2i(L-\phi)}\left(1 - N^2e^{-2\theta}\right)} + O\!\left(e^{-4\theta}\right) \qquad (19)$$

After including the propagation factors from regions 1 and 3 we obtain the transmission amplitude at resonance,

$$T(E) = e^{-i\pi/4}e^{-\theta}iNe^{-i\phi/2}\frac{e^{iL}}{1 - e^{2i(L-\phi)}\left(1 - N^2e^{-2\theta}\right)}Ne^{-i\phi/2}e^{-\theta}e^{i\pi/4}$$

$$= ie^{-i(L-\phi)} + O(e^{-4\theta}). \qquad (20)$$

In the second line we had invoked the resonance condition. Therefore, at resonance, the emergent beam is essentially identical to the incident beam, except for a phase. We have effectively obtained the transmission formula for both off-resonance and resonance cases correct up to terms of order $O(e^{-4\theta})$, which are generally very small. This is certainly much better than previous results using WKB [7, 14].

## III. Oscillating Potential Well

We examine now, within the previous formulation, the effect of an oscillating potential well in region 2 when the resonance condition holds. Unlike the previous situation this present one is a time-dependent problem. The new action $S[x]$ takes the form

$$S[x] = \int dt \left[\frac{1}{2}m\dot{x}^2 - V(x)\right] - \int dt\, V_1 \cos\omega t, \qquad (21)$$

in which the first integral gives the action for the time-independent problem while $V_{\text{well}}(t) = V_1 \cos\omega t$ represents the oscillating well. Expanding the action in terms of the classical path $x_{\text{cl}}$ for the time-independent potential, i.e., $x = x_{\text{cl}} + \delta x$, we have, in the semi-classical approximation,

$$D(x',t';x,t) = e^{(i/\hbar)S_0[x_{\text{cl}}]} \int D[\delta x]\, e^{(i/\hbar)\int dt\left[\frac{1}{2}m\delta\dot{x}^2 - V''[x_{\text{cl}}]\delta x^2\right]} \left(\sum_{n,m} J_n(f) J_m(f)\, e^{-in\omega t} e^{im\omega t'}\right) \qquad (22)$$

in which $f \equiv V_1/\hbar\omega$ and we had used the expansion in terms of Bessel functions,

$$e^{-if \sin\omega t} = \sum_{n=-\infty}^{\infty} J_n(f) e^{-in\omega t}. \qquad (23)$$

$S_0[x_{\text{cl}}]$ is the classical action without $V_{\text{well}}$. We choose the endpoints $x, x'$ to be sufficiently far away from the double barrier. Notice that all the time dependence is lumped into the expression inside parentheses. Therefore, we may apply the standard method of integrating the above and then go over to energy space by taking an inverse Fourier transform [12, 14]. All that is required is some care in interpreting the energy because the particle can exchange energy with the well. We write in place of Eq. (3)

$$\frac{m}{2\pi\hbar\sqrt{k(x')k(x)}}\exp(ik(y)dy)\sum_{l=-\infty}^{\infty}J_l(f)J_{m+l-n}(f)e^{i(E+n\omega-l\omega)t'}e^{-i(E+m\omega+l\omega)t} \qquad (24)$$

which we interpret as follows: the particle, which is initially incident from the left with the initial energy $E+\hbar\omega n$, loses $l$ photons of frequency $\omega$ upon entering the well and gains $m+l-n$ photons of frequency $\omega$ on leaving the well. It emerges with a final energy of $E+\hbar\omega m$.

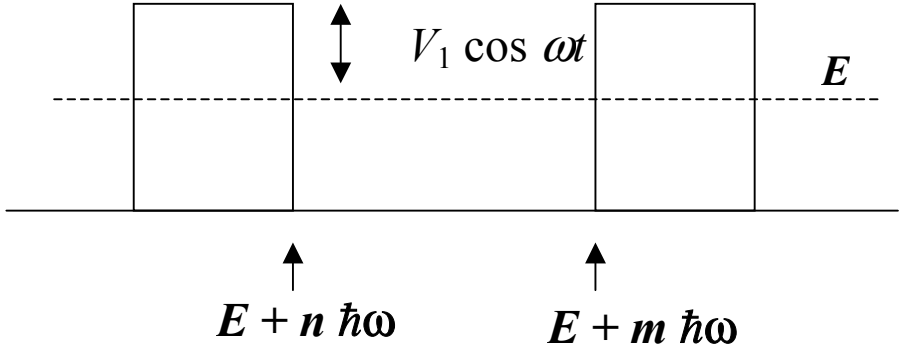

FIG. 2  Incident beam with energy $E+\hbar\omega n$ emerges with energy $E+\hbar\omega\, m$ after crossing the oscillating well.

Suppose that a monochromatic beam with initial energy $E+\hbar\omega n$ is incident upon the double well from the left (see Fig. 2). We would like to study the probability that the final emergent beam has the energy $E+\hbar\omega m$. Let the incident beam be given by $A^{(n)}e^{ikx}$ and the emergent beam by $B^{(m)}e^{ik'x}$. Then guided by Eq. (3) and the last result

$$B^{(m)} = \sum_{n,l} J_{n-l}(f) J_{m-l}(f) D^{(n,l)} A^{(n)}, \qquad (25)$$

the sum being from $-\infty$ to $+\infty$. Here $D^{(n,l)}$ is the propagator connecting the incident particle in the stationary state $E+\hbar\omega n$ to its final state $E+\hbar\omega m$ via intermediate states $E+\hbar\omega l$. Note that together with $l$, the sum above involves $n$ since many initial states can contribute to a definite final state. In principle the $D$'s also depend on $m$ but it is not necessary to explicitly display it here. At resonance $e^{-i(L-\phi)}=1$, and the presence of an oscillating well introduces small variations to the argument, so we have from Eq. (20), $D=ie^{-i(L-\phi)} \cong i[1-i(L-\phi)_1]$ to first order in $\hbar\omega$. Equation (25) can be cast as (we ignore an overall phase)

$$B^{(m)} = \sum_{n,l} J_{n-l}(f) J_{m-l}(f)[1-(L-\phi)_1] A^{(n)}. \qquad (26)$$

[Strictly speaking we should also expand the non-resonant term in the transmission amplitude; however, since it is proportional to sin $(L-\phi)$, its contribution can be lumped into the coefficient of $(L-\phi)_1$ in Eq. (26). This would not alter the final conclusion below.] Evaluating the correction we find

$$(L-\phi)_1 \cong \hbar\omega\{-(k'd+\sin\phi(k'/k))l+\sin\phi(q'/q)m+k'(d+k^{-1}\sin\phi)n\}. \qquad (27)$$

The primes denote differentiation with respect to the energy. Following Wagner, we try an ansatz of the form $A^{(n)} = J_n(\gamma f)$ where $\gamma$ is a parameter to be determined below (this also holds for $B^{(m)}$) [13]. Substituting into Eq. (26) and making use of the von Neumann summation formulas, namely [16],

$$\sum_{n,l} J_{n-l}(u) J_{m-l}(u) J_n(\gamma u) = J_m(\gamma u)$$

$$\sum_{n,l} l J_{n-l}(u) J_{m-l}(u) J_n(\gamma u) = m\left(1-\frac{1}{\gamma}\right) J_m(\gamma u) \qquad (28)$$

we have

$$B^{(m)} = \left(1 - i\hbar\omega m \frac{q'}{q}\sin\phi\right)A^{(m)} - \hbar\omega m(L+\sin\phi)\frac{k'}{k}\left(1-\frac{1}{\gamma}\right)A^{(m)} + \hbar\omega m \frac{k'}{k}(L+\sin\phi)A^{(m)}$$

(29)

To be consistent with the ansatz, we now solve for $\gamma$ under the condition that $A^{(m)} = B^{(m)}$. This gives

$$\gamma = -\frac{L+\sin\phi}{\frac{q'}{q}\frac{k}{k'}\sin\phi}$$

(30)

which is positive since the derivatives have opposite signs. Note that $\gamma$ depends only on the incoming kinetic energy and the barrier height and spacing. Our expression for $\gamma$ differs slightly from Wagner's, presumably on account of differences in his and our approximations [13].

Because the Bessel functions satisfy the normalization condition $\sum_n |J_n(u)|^2 = 1$, the probability for an electron to be in a sideband $n$ is $\int dx |\psi_n|^2 = |J_n(\gamma f)|^2$. For $V_1 = 0$ only the central band is present; as $V_1$ is increased more and more sidebands become important and the sideband probabilities oscillate. But the sidebands vanish completely at the zeros of $J_n(\gamma f)$. We have thus verified that for particular values of $f = V_1/\hbar\omega$ a simultaneous quenching of the resonant transmission probability occurs for all sidebands. Since, as we observed earlier, time-dependent problems are not usually studied within the semiclassical approach, this result shows that such problems may be addressed without much alteration of this formulation.

## IV. Conclusion

We had shown how the semiclassical approach can be employed to adequately describe the transmission probability for tunneling through a double barrier structure both at resonance and off resonance if two sets of amplitude factors and phases are used. We had also shown how this method can be formulated to address a time-dependent version of the same structure; the novel feature of 'coherent destruction of tunneling' may be studied through this approach. Although we carried out an expansion to first order in $\hbar\omega$ our result is not perturbative in the modulation amplitude $V_1$. Our hope is to extend this work to general barrier penetration problems that are not readily amenable to an analytical solution.

**Figures & captions**

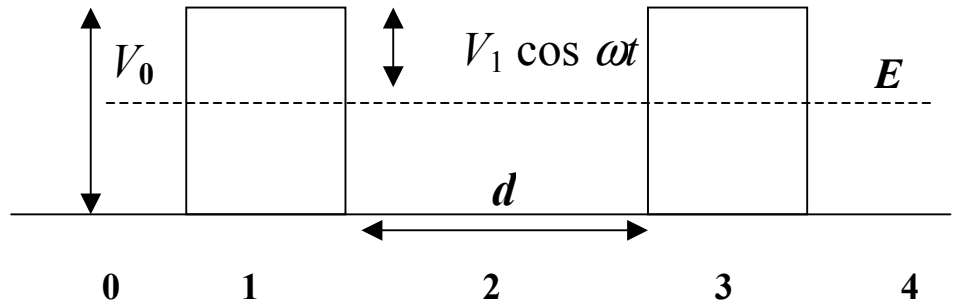

**FIG 1 Transmission through a double barrier structure with an oscillating quantum well between the static barriers.**

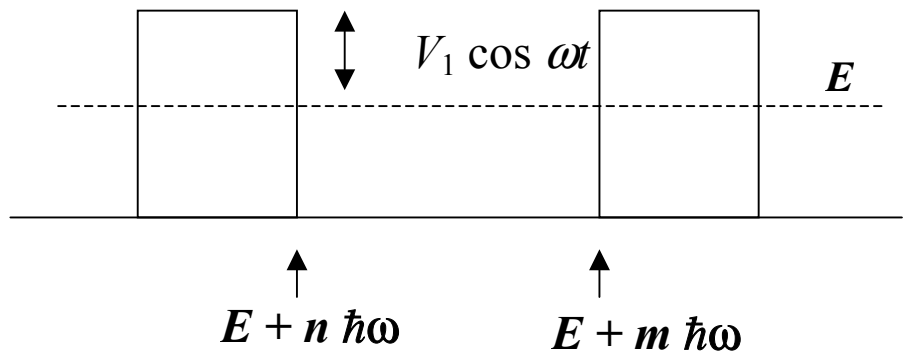

**FIG. 2 Incident beam with energy $E+\hbar\omega n$ emerges with energy $E+\hbar\omega m$ after crossing the oscillating well.**

**Table**

Table I: Phase and weight factors at each turning point.

|  | Phase | Weight |
|---|---|---|
| Allowed → allowed | $e^{-i\phi}$ | 1 |
| Forbidden→forbidden | $-e^{-i(\phi-\bar{\phi})/2}$ | $N/2\bar{N}$ |
| Forbidden→allowed | $ie^{-i\phi/2}$ | $N$ |
| Allowed→forbidden | $e^{-i\phi/2}$ | $N$ |